\documentclass[twocolumn,aps,prl,amsmath,amssymb]{revtex4-1}
\usepackage[latin9]{inputenc}
\usepackage{amstext}
\usepackage{amssymb}
\usepackage{graphicx}
\usepackage{esint}

\makeatletter
\usepackage{amssymb}
\usepackage{mathbbold}
\usepackage{amsfonts}
\usepackage{epsfig}
\usepackage{bm}\usepackage{dcolumn}\usepackage{color}
\usepackage{overpic}
\usepackage[bottom]{footmisc}

\makeatother

\begin{document}

\title{Realizing Fulde-Ferrell Superfluids via a Dark-State Control of Feshbach Resonances}

\author{Lianyi He$^{1}$} 
\email{lianyi@mail.tsinghua.edu.cn}

\author{Hui Hu$^{2}$}
\email{hhu@swin.edu.au}

\author{Xia-Ji Liu$^{2}$}
\email{xiajiliu@swin.edu.au}

\affiliation{$^{1}$ Department of Physics and State Key Laboratory of Low-Dimensional
Quantum Physics, Tsinghua University, Beijing 100084, China}
\affiliation{ $^{2}$ Centre for Quantum and Optical Science, Swinburne University
of Technology, Melbourne 3122, Australia}

\date{\today}
\begin{abstract}
We propose that the long-sought Fulde-Ferrell superfluidity with nonzero 
momentum pairing can be realized in ultracold two-component Fermi
gases of $^{40}$K or $^{6}$Li atoms by optically tuning their magnetic
Feshbach resonances via the creation of a closed-channel dark state
with a Doppler-shifted Stark effect. In this scheme, two counterpropagating
optical fields are applied to couple two molecular states in the closed
channel to an excited molecular state, leading to a significant violation
of Galilean invariance in the dark-state regime and hence to the possibility
of Fulde-Ferrell superfluidity. We develop a field theoretical formulation
for both two-body and many-body problems and predict that the Fulde-Ferrell
state has remarkable properties, such as anisotropic single-particle
dispersion relation, suppressed superfluid density at zero temperature,
anisotropic sound velocity and rotonic collective mode. The latter
two features can be experimentally probed using Bragg spectroscopy,
providing a smoking-gun proof of Fulde-Ferrell superfluidity.
\end{abstract}


\maketitle
\emph{Introduction.} The application of magnetic Feshbach resonance
(MFR) in Fermi gases of alkali-metal atoms \cite{Chin2010}, i.e.,
tuning the interatomic interaction strength, opens a new paradigm
to study strongly correlated many-body phenomena \cite{Bloch2008,Giorgini2008}.
The crossover from Bardeen-Cooper-Schrieffer (BCS) superfluid to Bose-Einstein
condensate (BEC) \cite{Eagles1969,Leggett1980,Nozieres1985,Melo1993,Chen2005,Gurarie2007}
in atomic Fermi gases has now been experimentally explored in great
detail \cite{Greiner2003,Jochim2003,Zwierlein2003,Nascimbene2010,Horikoshi2010,Ku2012},
leading to a number of new concepts such as unitary Fermi superfluid
and universal equation of state \cite{Ku2012,Ho2004,Hu2007-1} that
bring new insights to better understand other strongly interacting
systems in nature \cite{Lee2006RMP,Lee2006PRC,Kolb2004}.

Finite-momentum pairing superfluidity, or the so-called Fulde-Ferrell-Larkin-Ovchinikov
(FFLO) state \cite{Fulde1964,Larkin1964}, is another intriguing phenomenon
addressed using ultracold Fermi gases near MFR \cite{Zwierlein2006,Partridge2006,Sheehy2006,Sheehy2007,Hu2007-2,Orso2007}.
It has been studied and pursued for over a half-century in both condensed
matter physics and nuclear physics \cite{Casalbuoni2004,Anglani2014}.
Yet, its existence remains elusive. In three-dimensional free space, the conventional
scenario of spin-population imbalance leads to a rather narrow window
for FFLO in atomic Fermi gases \cite{Sheehy2006,Sheehy2007}. It was proposed that the stability regime for FFLO can be significantly 
enhanced via engineering single-particle properties \cite{Torma2017}, using optical lattice \cite{Torma2007,Torma2008,Torma2010,Trivedi2010,Trivedi2011, Tempere2011-1,Tempere2011-2,Tempere2011-3} or spin-orbit coupling \cite{Dong2013-1,Zheng2013,Wu2013,Liu2013,Hu2013,Dong2013-2,Iskin2013,Shenoy2013,Zhou2014}.
It was theoretically shown that in the presence of spin-orbit coupling, the Fulde-Ferrell (FF) superfluid
state is energetically favored in a large parameter space because
of the violation of Galilean invariance, which sets a preferable momentum
for Cooper pairs in the presence of an in-plane Zeeman field \cite{Zheng2013,Wu2013,Liu2013,Hu2013}.
However, the heating problem in realizing spin-orbit coupled FF superfluids at low temperature has not yet been solved experimentally
\cite{Zhai2015}.

%%%%%%%%%%%%%%%%%%%%%%%%%%%%%%%%%%%%%%%%%%%%%%%%%%%%%%%%%%%%%%%%%%%%%%%
\begin{figure}[!tbh]
\centering{}\includegraphics[width=0.45\textwidth]{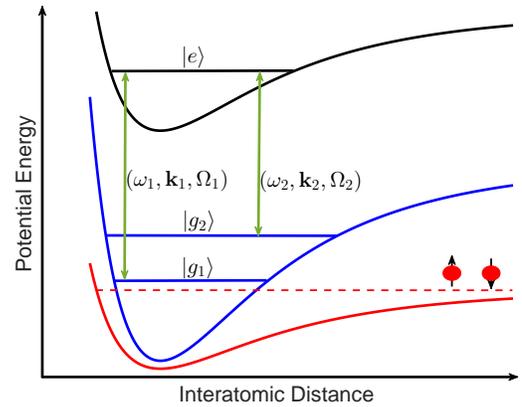} \caption{(Color online) Level scheme for the dark-state optical control of
MFR. The ground molecular state $\left|g_{1}\right\rangle $, responsible
for a magnetic Feshbach resonance, is shifted by two optical fields
(green lines). The small Doppler effect in such a Stark shift due
to ${\bf k}_{1}\protect\neq{\bf k}_{2}$ can be greatly amplified
in the dark-state regime by a factor of $(\Omega_{1}/\Omega_{2})^{2}\sim10-10^{3}$,
leading to a significant violation of Galilean invariance. \label{fig1}}
\end{figure}

%%%%%%%%%%%%%%%%%%%%%%%%%%%%%%%%%%%%%%%%%%%%%%%%%%%%%%%%%%%%%%%%%%%%%%%%

In this Letter, we propose that the Fulde-Ferrell superfluidity can be realized without spin-population imbalance, via engineering interactomic interaction. The new scenario is based on the recent
ground-breaking demonstration of a dark-state optical control of MFRs
\cite{Jagannathan2016} and its innovative extension to allow a center-of-mass (CoM)
momentum ${\bf q}$-dependent interatomic interaction \cite{Jie2016}.
Here, the MFR is induced by the hyperfine coupling between the atomic
pair state in the open channel and a molecular state $|g_{1}\rangle$
in the closed channel \cite{Timmermans1999,Holland2001,Ohashi2003}.
As shown in Fig. \ref{fig1}, the dark-state optical control of the
MFR uses two ground molecular states $|g_{1}\rangle$ and $|g_{2}\rangle$
that are coupled to an excited molecular state $|e\rangle$ by two
optical fields of frequencies $\omega_{1}$ and $\omega_{2}$, wave
vectors ${\bf k}_{1}$ and ${\bf k}_{2}$, and Rabi frequencies $\Omega_{1}$
and $\Omega_{2}$, respectively \cite{Jagannathan2016,Wu2012-1,Wu2012-2}.
In the dark-state regime, the resulting Stark shift $\Sigma_{1}$
in the state $|g_{1}\rangle$ is affected by the Doppler effect \cite{Jie2016},
i.e., $\Sigma_{1}\sim(\Omega_{1}/\Omega_{2})^{2}{\bf q}\cdot({\bf k}_{1}-{\bf k}_{2})$,
which breaks the Galilean invariance of the system. Hence, if the
two optical fields propagate along opposite directions (i.e., ${\bf k}_{1}=-{\bf k}_{2}=k_{\rm R}\mathbf{e}_{z}$),
the violation of Galilean invariance becomes significant when $(\Omega_{1}/\Omega_{2})^2\gg1$,
and may lead to interesting many-body consequences.

One of the key observations in this Letter is that the zero-momentum
pairing state has a nonzero current ${\bf j}\propto{\bf k}_{1}-{\bf k}_{2}$
carried by the condensate and suffers from severe instability. The
true ground state of the system therefore falls toward a FF state so that
the currents carried by the condensate and the fermionic quasiparticles
cancel each other precisely. This compensation mechanism is equally
important for reducing the Doppler effect in the two-photon detuning
and keeps the system in the dark-state regime. As a result, optical
loss is negligible and the Fermi cloud does not suffer from heating
as in the case of spin-orbit coupling. We predict that the FF state
realized by our proposal has some unique features, including the anisotropic
phonon dispersion and the emergence of a roton structure in the collective
modes, both of which can be readily examined in cold-atom experiments
as clear evidences of the long-sought FF superfluidity. 

\emph{Field theory.} We start by formulating a field theoretical description of the optical control of MFR, which
provides a convenient way to perform many-body calculations. In the
absence of optical fields, the MFR can be described by the atom-molecule
theory \cite{Gurarie2007,Timmermans1999,Holland2001,Ohashi2003}.
The Lagrangian density is given by ${\cal L}_{{\rm MFR}}={\cal L}_{{\rm A}}+{\cal L}_{{\rm M}}+{\cal L}_{{\rm AM}}$,
with 
\begin{eqnarray}
{\cal L}_{{\rm A}} & = & \sum_{\sigma=\uparrow,\downarrow}\psi_{\sigma}^{\dagger}\hat{K}_{{\rm F}}\psi_{\sigma}^{\phantom{\dag}}-u_{0}\psi_{\uparrow}^{\dagger}\psi_{\downarrow}^{\dagger}\psi_{\downarrow}^{\phantom{\dag}}\psi_{\uparrow}^{\phantom{\dag}},\nonumber \\
{\cal L}_{{\rm M}} & = & \varphi_{1}^{\dagger}\left(\hat{K}_{{\rm B}}-\nu_{0}\right)\varphi_{1}^{\phantom{\dag}},\nonumber \\
{\cal L}_{{\rm AM}} & = & -g_{0}\left(\varphi_{1}^{\dagger}\psi_{\downarrow}^{\phantom{\dag}}\psi_{\uparrow}^{\phantom{\dag}}+\varphi_{1}^{\phantom{\dag}}\psi_{\uparrow}^{\dagger}\psi_{\downarrow}^{\dagger}\right).
\end{eqnarray}
Here $\psi_{\sigma}$ ($\sigma=\uparrow,\downarrow$) denotes the
open-channel fermions and $\varphi_{1}$ denotes the closed-channel
molecular state $|g_{1}\rangle$. We use the notations $\hat{K}_{{\rm F}}=i\partial_{t}+\nabla^{2}/(2m)$
and $\hat{K}_{{\rm B}}=i\partial_{t}+\nabla^{2}/(4m)$ with $t$ being
the time and $m$ being the atom mass. The units $\hbar=k_{{\rm B}}=1$
will be used throughout. The bare couplings $u_{0}$ and $g_{0}$
as well as the bare magnetic detuning $\nu_{0}$ should be renormalized
in terms of the background scattering length $a_{{\rm bg}}$, resonance
width $\Delta B$, and detuning $\Delta\mu(B-B_{0})$, in the forms
of $u=4\pi a_{{\rm bg}}/m$, $g=\sqrt{\Delta\mu\Delta B u}$,
and $\nu=\Delta\mu(B-B_{0})$ \cite{Timmermans1999,Holland2001,Ohashi2003,SUPP}.
In the presence of optical fields, we add a new molecular part 
\begin{eqnarray}\label{L-Mprime}
{\cal L}_{{\rm M}}^{\prime} & = & \varphi_{2}^{\dagger}\left(\hat{K}_{{\rm B}}-E_{2}\right)\varphi_{2}^{\phantom{\dag}}+\varphi_{{\rm e}}^{\dagger}\left(\hat{K}_{{\rm B}}-E_{{\rm e}}+i\frac{\gamma_{{\rm e}}}{2}\right)\varphi_{{\rm e}}^{\phantom{\dag}}\nonumber \\
 & - & \sum_{l=1,2}\left[\frac{\Omega_{l}}{2}\varphi_{l}^{\phantom{\dag}}\varphi_{{\rm e}}^{\dagger}e^{i\theta_{l}({\bf r},t)}+\frac{\Omega_{l}^{*}}{2}\varphi_{l}^{\dagger}\varphi_{{\rm e}}^{\phantom{\dag}}e^{-i\theta_{l}({\bf r},t)}\right],
\end{eqnarray}
where $\varphi_{2}$ and $\varphi_{{\rm e}}$ denote the states $|g_{2}\rangle$
and $|e\rangle$ with energies $E_{2}$ and $E_{{\rm e}}$, respectively,
and $\theta_{l}({\bf r},t)={\bf k}_{l}\cdot{\bf r}-\omega_{l}t$.
The spontaneous decay of the excited molecular state $|e\rangle$
is treated phenomenologically by a decay rate $\gamma_{{\rm e}}$. The last term in Eq. (\ref{L-Mprime}) describes the one-body Raman transitions 
between the molecular states.

The phase factors $\theta_{l}({\bf r},t)$ can be eliminated by defining
two new molecular fields, $\phi_{{\rm e}}=\varphi_{{\rm e}}e^{-i\theta_{1}}$
and $\phi_{2}=\varphi_{2}e^{-i(\theta_{1}-\theta_{2})}$. By setting
$\phi_{1}=\varphi_{1}$, we can express the molecular part in a compact
form ${\cal L}_{{\rm M}}^{\phantom{\dag}}+{\cal L}_{{\rm M}}^{\prime}=\Phi^{\dagger}{\bf M}(i\partial_{t},-i\nabla)\Phi$,
where $\Phi=(\phi_{1},\ \phi_{2},\ \phi_{{\rm e}})^{{\rm T}}$ and
the inverse propagator matrix in momentum space reads 
\begin{eqnarray}
{\bf M}(q_{0},{\bf q})=\left(\begin{array}{ccc}
I_{1}(q_{0},{\bf q}) & 0 & -\Omega_{1}^{*}/2\\
0 & I_{2}(q_{0},{\bf q}) & -\Omega_{2}^{*}/2\\
-\Omega_{1}/2 & -\Omega_{2}/2 & I_{{\rm e}}(q_{0},{\bf q})
\end{array}\right),
\end{eqnarray}
with diagonal elements $I_{1}(q_{0},{\bf q})=Z-\nu_{0}$ and 
\begin{eqnarray}
 &  & I_{2}(q_{0},{\bf q})=Z-\frac{\mathbf{q}\cdot({\bf k}_{1}-{\bf k}_{2})}{2m}-\frac{({\bf k}_{1}-{\bf k}_{2})^{2}}{4m}+\delta,\nonumber \\
 &  & I_{{\rm e}}(q_{0},{\bf q})=Z-\frac{{\bf q\cdot k}_{1}}{2m}-\frac{{\bf k}_{1}^{2}}{4m}+\Delta_{{\rm e}}+i\frac{\gamma_{{\rm e}}}{2}.\label{eq:I2Ie}
\end{eqnarray}
Here, $\Delta_{{\rm e}}=\omega_{1}-E_{{\rm e}}$ is the one-photon
detuning, $\delta=(\omega_{1}-\omega_{2})-E_{2}$ is the two-photon detuning,
and $Z=q_{0}-{\bf q}^{2}/(4m)$ is a Galilean invariant combination, with $q_0$ and ${\bf q}$ being the CoM energy and momentum of two incident atoms. 
The Rabi frequencies $\Omega_1$ and $\Omega_2$ as well as the detunings $\Delta_{\rm e}$ and $\delta$ are experimentally tunable \cite{Jagannathan2016,Fu2013}.

\emph{Two-body problem.} To solve the two-body problem, we compute the off-shell $T$-matrix
for atom-atom scattering, $T_{{\rm 2b}}(q_{0},{\bf q})=[U^{-1}(q_{0},{\bf q})-\Pi(q_{0},{\bf q})]^{-1}$,
which is exactly given by the bubble summation. Here, $U(q_{0},{\bf q})=u_{0}+g_{0}^{2}D_{1}(q_{0},{\bf q})$
is an energy- and momentum-dependent interaction vertex, with $D_{1}(q_{0},{\bf q})$
being the propagator of the molecular state $|g_{1}\rangle$. With
optical fields, $D_{1}(q_{0},{\bf q})=[I_{1}(q_{0},{\bf q})-\Sigma_{1}(q_{0},{\bf q})]^{-1}$
is given by the 11-component of ${\bf M}^{-1}(q_{0},{\bf q})$, where
the self-energy or the so-called Stark shift reads 
\begin{eqnarray}
\Sigma_{1}(q_{0},{\bf q})=\frac{|\Omega_{1}|^{2}}{4}\left[I_{{\rm e}}(q_{0},{\bf q})-\frac{|\Omega_{2}|^{2}}{4I_{2}(q_{0},{\bf q})}\right]^{-1}.\label{eq: StarkShift}
\end{eqnarray}
The two-atom bubble function $\Pi(q_{0},{\bf q})$ is given by $\Pi(q_{0},{\bf q})=\sum_{{\bf p}}(Z+i0^{+}-2\varepsilon_{{\bf p}})^{-1}$
with $\varepsilon_{{\bf p}}={\bf p}^{2}/(2m)$, and is to be replaced
by $\Pi_{{\rm R}}(q_{0},{\bf q})=[m/(4\pi)]\sqrt{-m(Z+i0^{+})}$ after
renormalization. More explicitly, in terms of the renormalized quantities,
the $T$-matrix $T_{{\rm 2b}}(q_{0},{\bf q})$ takes the form, $T_{{\rm 2b}}(q_{0},{\bf q})=[U_{\rm R}^{-1}(q_{0},{\bf q})-\Pi_{\rm R}(q_{0},{\bf q})]^{-1}$,
where the effective coupling reads \cite{SUPP}
\begin{eqnarray}
U_{{\rm R}}(q_{0},{\bf q})=u+\frac{g^{2}}{Z-\nu-\Sigma_{1}(q_{0},{\bf q})},\label{LM-coupling}
\end{eqnarray}
which fully characterizes the interatomic interaction in the presence
of laser beams.

For the optical control of MFRs in atomic gases of $^{6}$Li and $^{40}$K,
the Doppler effect to the Stark shift, i.e., the term $\mathbf{q}\cdot({\bf k}_{1}-{\bf k}_{2})/(2m)$
in Eqs. (\ref{eq:I2Ie}) and (\ref{eq: StarkShift}), is of the order
of the recoil energy $E_{\rm R}=k_{\rm R}^{2}/(2m)\sim2\pi\times10$ kHz and
is usually neglected, in comparison with the decay rate and Rabi frequencies
$\gamma_{{\rm e}},\Omega_{1,2}\sim2\pi\times10$ MHz. However, in
the dark-state regime with $\delta=0$ (i.e., $I_{\rm e}\ll\Omega_{2}^{2}/I_{2}$)
and a large ratio $\Omega_{1}/\Omega_{2}$, it could be greatly
enhanced, leading to a Stark shift as large as $10^{-2}\Delta\mu\Delta B$. This gives rise to a CoM momentum dependent interaction
\cite{Jie2016} and hence a strong violation of Galilean invariance.
Throughout the work, we assume ${\bf k}_{1}=k_{\rm R}\mathbf{e}_{z}=-{\bf k}_{2}$
with $k_{\rm R}=8.138\times10^{6}$ m$^{-1}$ and focus on the case of
$^{40}$K atoms near the broad resonance at $B_{0}=202.02$ G with
$a_{\textrm{bg}}=174a_{0}$, $\Delta B=7.04$ G, and $\Delta\mu=2\mu_{\rm B}$
\cite{Gaebler2010}. We consider the typical values $\Delta_{{\rm e}}=-2\pi\times500$
MHz, $\gamma_{{\rm e}}=2\pi\times6$ MHz, $\delta=0$, $\Omega_{1}=2\pi\times120$
MHz and $\Omega_{2}=2\pi\times20$ MHz, unless specified elsewhere
\cite{Fu2013}. We also take a typical atom density $n=1.82\times10^{13}$cm$^{-3}$,
corresponding to a Fermi momentum $k_{{\rm F}}=(3\pi^{2}n)^{1/3}$$\simeq k_{\rm R}$
\cite{Fu2013}. 

%%%%%%%%%%%%%%%%%%%%%%%%%%%%%%%%%%%%%%%%%%%%%%%%%%%%%%%%%%%%%%%%%%%%%%%
\begin{figure}
\centering{}\includegraphics[width=0.45\textwidth]{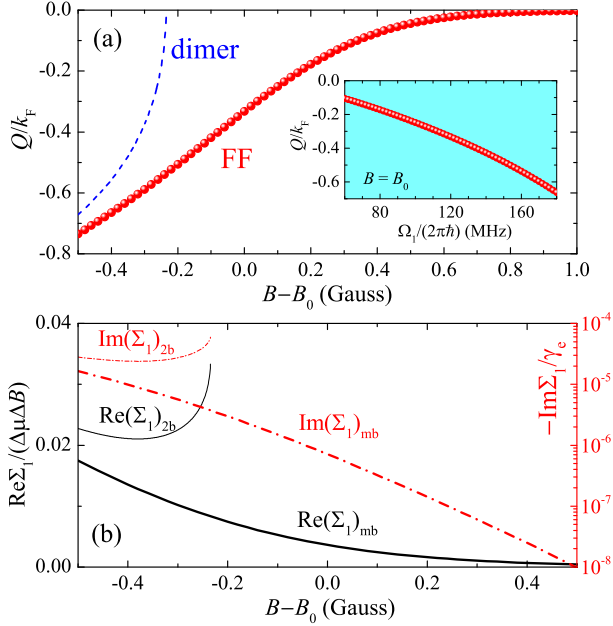} \caption{(Color online) (a) The momentum of the dimer bound state and the pairing
momentum of the FF superfluid as a function of the magnetic detuning
$B-B_{0}$. The inset shows the dependence of the FF momentum on $\Omega_{1}$
at resonance $B=B_{0}$. (b) Two- and many-body Stark shifts in
the BEC-BCS crossover. We take $(q_{0},{\bf q})=(E_{\rm d}(\mathbf{q}_{\rm d}),{\bf q}_{\rm d})$
and $(q_{0},{\bf q})=(2\mu,{\bf Q})$ for the two- and many-body cases, respectively.
\label{fig2}}
\end{figure}

%%%%%%%%%%%%%%%%%%%%%%%%%%%%%%%%%%%%%%%%%%%%%%%%%%%%%%%%%%%%%%%%%%%%%%%%

With the above parameters, the violation of Galilean invariance is
already clearly seen in the dimer bound state below the MFR, whose
energy $E_{{\rm d}}({\bf q})$ is determined by the pole of the $T$-matrix,
i.e., $T_{{\rm 2b}}^{-1}[E_{{\rm d}}({\bf q}),{\bf q}]=0$ \cite{SUPP,Note-gammae}.
Without optical fields, the Galilean invariance ensures that $E_{{\rm d}}({\bf q})=\varepsilon_{{\rm B}}+{\bf q}^{2}/(4m)$,
with $\varepsilon_{{\rm B}}$ being the binding energy, and the dimer
has lowest energy at ${\bf q}=0$. In the presence of optical fields,
it is obvious that the effective interaction $U_{{\rm R}}(q_{0},{\bf q})$
depend not only on $Z$ but also on the pair momentum ${\bf q}$ itself,
which indicates that the Galilean invariance and especially the spatial
inversion symmetry are broken. As a consequence, $E_{{\rm d}}({\bf q})$
has a nontrivial ${\bf q}$ dependence and the lowest dimer energy
locates at ${\bf q}\neq0$. In Fig. \ref{fig2}(a), we show the momentum
of the dimer bound state, ${\bf q}_{{\rm d}}=Q{\bf e}_{z}$, by using
a dashed line. We have in general $Q\neq0$ at the BEC side of the
MFR. The corresponding two-body Stark shift is reported in Fig. \ref{fig2}(b).
Its imaginary part (i.e., decay rate) is about $10^{-5}\gamma_{\textrm{e}}\sim2\pi\times100$
Hz, indicating a reasonably long dimer lifetime $\sim0.01-0.1$ s
\cite{Jie2016, SUPP}.

\emph{Many-body theory.} The partition function of the system is given
by the imaginary-time formalism $\mathcal{Z}=\int\mathcal{D}[\psi,\psi^{\dagger};\Phi,\Phi^{\dagger}]\exp[\int dx({\cal L_{\textrm{MFR}}}+{\cal L}_{\rm M}^\prime+{\cal L}_{\mu})]$,
where $x=(\tau,{\bf r})$ and the chemical potential $\mu$ is introduced
through the term ${\cal L}_{\mu}=\mu\sum_{\sigma=\uparrow,\downarrow}\psi_{\sigma}^{\dagger}\psi_{\sigma}^{\phantom{\dag}}+2\mu\sum_{l=1,2,{\rm e}}\phi_{l}^{\dagger}\phi_{l}^{\phantom{\dag}}$.
To decouple the four-fermion interaction term, we introduce an auxiliary
field $\phi_{{\rm f}}(x)=u_{0}\psi_{\downarrow}(x)\psi_{\uparrow}(x)$,
perform the Hubbard-Stratonovich transformation, and integrate out
the fermions to obtain ${\cal Z}=\int\mathcal{D}[\phi_{\textrm{f}}^{\phantom{\dag}},\phi_{\textrm{f}}^{\dagger};\Phi,\Phi^{\dagger}]\exp{\left(-{\cal S}_{{\rm eff}}\right)}$,
with the effective action [$\Delta(x)=\phi_{{\rm f}}(x)+g_{0}\phi_{1}(x)$],
\begin{eqnarray}
{\cal S}_{{\rm eff}} & = & -\ {\rm Tr}\ln\left[\left(\begin{array}{cc}
\hat{K}_{{\rm F}}+\mu & \Delta(x)\\
\Delta^{\dagger}(x) & -\hat{K}_{{\rm F}}^{*}-\mu
\end{array}\right)\delta(x-x^{\prime})\right]\nonumber \\
 &  & -\int dx\ \left[\frac{|\phi_{{\rm f}}(x)|^{2}}{u_{0}}+\Phi^{\dagger}{\bf M}(2\mu-\partial_{\tau},-i\nabla)\Phi\right].
\end{eqnarray}

We evaluate $\mathcal{Z}$ in the mean-field approximation, which
amounts to searching for the static saddle-point solution $\phi_{l}(x)=\bar{\phi}_{l}({\bf r})$
($l=1,2,{\rm e},{\rm f}$) that minimizes the effective action ${\cal S}_{{\rm eff}}$
(i.e., $\delta{\cal S}_{{\rm eff}}/\delta\bar{\phi}_{l}^{\phantom{\dag}}({\bf r})=0$
and $\delta{\cal S}_{{\rm eff}}/\delta\bar{\phi}_{l}^{*}({\bf r})=0$).
Motivated by the fact that the dimer ground state has nonzero momentum,
we expect that the fermion pairing favors nonzero momentum in the
superfluid state. Thus, we take the Fulde-Ferrell ansatz for the saddle-point
solution, $\bar{\phi}_{l}({\bf r})=C_{l}e^{i{\bf Q}\cdot{\bf r}}$,
where ${\bf Q}$ is the pairing momentum. The fermionic part (i.e.,
the ${\rm Tr}\ln$ term) can be evaluated by performing a phase transformation
for the fermion fields, $\psi_{\sigma}=\tilde{\psi}_{\sigma}e^{i{\bf Q}\cdot{\bf r}/2}$.
Using the saddle-point condition $\partial{\cal S}_{{\rm eff}}/\partial C_{l}=0$,
we can express $C_{l}$ in terms of $\Delta=C_{{\rm f}}+g_{0}C_{1}$.
By further using the renormalized couplings and detuning, the thermodynamic
potential at $T=0$ reads \cite{SUPP,Note-gammae}
\begin{equation}
\Omega=\Omega_{\rm q}+\sum_{{\bf k}}\left(\xi_{{\bf k}}-E_{{\bf k}}+\frac{|\Delta|^{2}}{2\varepsilon_{{\bf k}}}\right)-\frac{|\Delta|^{2}}{U_{{\rm R}}(2\mu,{\bf Q})}.
\end{equation}
Here the dispersions are defined as $\xi_{{\bf k}}=\varepsilon_{{\bf k}}+{\bf Q}^{2}/(8m)-\mu$ and
$E_{{\bf k}}=(\xi_{{\bf k}}^{2}+|\Delta|^{2})^{1/2}$. The quasiparticle term $\Omega_{\rm q}=\sum_{s=\pm}\sum_{\bf k}E_{{\bf k}}^{s}\Theta(-E_{{\bf k}}^{s})$ contributes only when the quasiparticle exitations
$E_{{\bf k}}^{\pm}=E_{{\bf k}}\pm{\bf k}\cdot{\bf Q}/(2m)$ are gapless.
The last term in the expression is quite meaningful: The condensation
energy contains the effective two-body interaction $U_{{\rm R}}(q_{0},{\bf q})$
evaluated at $(q_{0},{\bf q})=(2\mu,{\bf Q})$. The superfluid
ground state is fully determined by the gap equations \cite{SUPP}:
$\partial\Omega/\partial\Delta=0$ and $\partial\Omega/\partial{\bf Q}=0$,
which minimize the thermodynamic potential in the energy landscape
spanned by $\Delta$ and ${\bf Q}$. The chemical potential is determined
by using the number equation, $n=-\partial\Omega/\partial\mu$. 

%%%%%%%%%%%%%%%%%%%%%%%%%%%%%%%%%%%%%%%%%%%%%%%%%%%%%%%%%%%%%%%%%%%%%%%
\begin{figure}
\centering{}\includegraphics[width=0.48\textwidth]{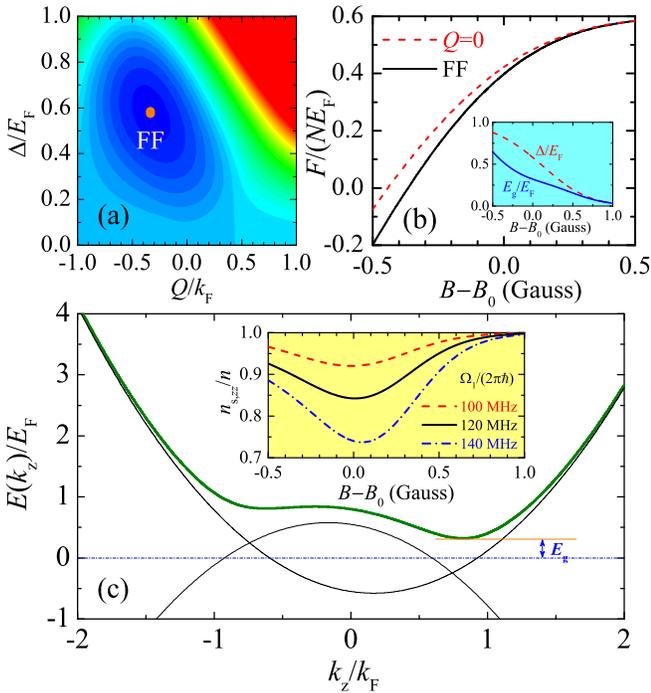} \caption{(Color online) (a) A contour plot of the thermodynamic potential in
the plane of $\Delta$ and $Q$, at $\mu\simeq0.58E_{{\rm F}}$ with
$E_{{\rm F}}=k_{{\rm F}}^{2}/(2m)$ and at resonance $B=B_{0}$, from
minimum (blue) to maximum (red). The FF state is highlighted by the
orange dot. (b) The free energy gain of the FF state, in comparison
with the BCS state with $Q=0$, as a function of the detuning $B-B_{0}$.
The latter is obtained by forcing $Q=0$. The inset reports the energy
gap and pairing gap. (c) The single-particle energy spectrum of the
FF state along the $k_{z}$ direction at $B=B_{0}$. The two thin
lines show the free-particle and free-hole energies, i.e., $\xi_{\mathbf{k}+\mathbf{Q}/2}$
and $-\xi_{-\mathbf{k}+\mathbf{Q}/2}$, respectively. The inset shows
the superfluid fraction along the $z$ direction at different Rabi
frequencies $\Omega_{1}$. \label{fig3}}
\end{figure}

%%%%%%%%%%%%%%%%%%%%%%%%%%%%%%%%%%%%%%%%%%%%%%%%%%%%%%%%%%%%%%%%%%%%%%%%

%%%%%%%%%%%%%%%%%%%%%%%%%%%%%%%%%%%%%%%%%%%%%%%%%%%%%%%%%%%%%%%%%%%%%%%
\begin{figure}
\centering{}\includegraphics[width=0.45\textwidth]{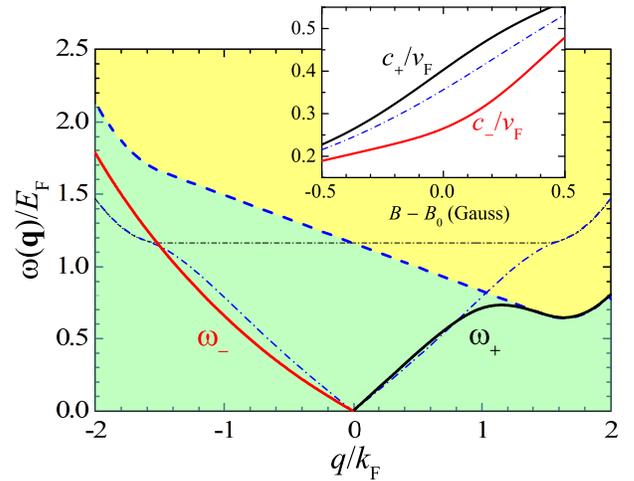} \caption{(Color online) The frequency of collective phonon modes at resonance
$B=B_{0}$ within the two-particle excitation gap, when the mode momentum
$\mathbf{q}$ is in the same ($\omega_{+}$) or opposite ($\omega_{-}$)
direction as the FF momentum $\mathbf{Q}$. The different mode frequencies
lead to two sound velocities, as shown in the inset, as a function
of $B-B_{0}$. The yellow area above the blue dashed line is the two-particle
continuum. The dot-dashed lines show the results when the mode momentum
$\mathbf{q}$ is perpendicular to the FF momentum, i.e., $\mathbf{q}\cdot\mathbf{Q}=0$.
\label{fig4}}
\end{figure}

%%%%%%%%%%%%%%%%%%%%%%%%%%%%%%%%%%%%%%%%%%%%%%%%%%%%%%%%%%%%%%%%%%%%%%%%

\emph{Finite-momentum superfluidity.} Before we show the mean-field
results, we present some analytical arguments which indicates that
the FF state is quite robust here. First, in the conventional FF problem
with Galilean invariance, the thermodynamic potential is an even function
of ${\bf Q}$ and gives a trivial solution ${\bf Q}=0$, which indicates
that the instability toward FF occurs at the order $O({\bf Q}^{2})$
\cite{He2006-1,He2006-2}. The scenario of mismatched Fermi surfaces
leads to a rather narrow window for FFLO. However, here we find that
${\bf Q}=0$ is no longer a trivial solution. Physically, this means
that the ${\bf Q}=0$ state has a spontaneously generated current
${\bf j}\neq0$ from the condensate due to the violation of Galilean
invariance, where ${\bf j}$ can be obtained by ${\bf j}=2m\partial\Omega/\partial{\bf Q}$
evaluated at ${\bf Q}=0$. Explicitly, we find ${\bf j}\propto {\bf k}_1-{\bf k}_2=2k_{{\rm R}}{\bf e}_{z}$
\cite{SUPP}. Thus the instability toward FF occurs at the order $O({\bf Q})$.
Therefore, to stabilize the system, the ground state falls to a FF
state so that a new current generated by the fermionic quasiparticles,
${\bf j}^{\prime}\propto{\bf Q}$, cancels precisely the current carried
by the condensate. This also shows that the pair momentum is along
the $z$ direction, ${\bf Q}=Q{\bf e}_{z}$.

On the other hand, in the BEC limit, $\mu$ becomes large and negative
and $|\mu|\gg\Delta$. To the leading order in $\Delta/|\mu|$, the
gap equation $\partial\Omega/\partial\Delta=0$ can be expressed as
\cite{SUPP} 
\begin{eqnarray}
U_{{\rm R}}^{-1}(2\mu,{\bf Q})-\Pi_{{\rm R}}(2\mu,{\bf Q})=0,
\end{eqnarray}
which is exactly the equation determining the dimer energy $E_{{\rm d}}({\bf Q})=2\mu({\bf Q})$
as a function of ${\bf Q}$. Moreover, using the fact $|\mu|\gg\Delta$,
we can show that the other two equations, $\partial\Omega/\partial{\bf Q}=0$
and $n=-\partial\Omega/\partial\mu$, give rise to the equation $\partial\mu({\bf Q})/\partial{\bf Q}=0$
\cite{SUPP}. Thus, $2\mu$ approaches the lowest energy of the dimer
state, located at finite momentum. Therefore, in the BEC limit, the
superfluid ground state is a finite-momentum Bose-Einstein condensation
of tightly bound dimers.

Fig. \ref{fig2}(a) reports a typical calculation of the FF momentum
$Q$ across the MFR (solid circles). We find that unlike the two-body
case (dashed line), the FF state with $Q\neq0$ arises even at the
BCS side. It is remarkable that the imaginary part of the many-body Stark
shift is very small (i.e., $<10^{-6}\gamma_{\textrm{e}}$) at the
BCS side {[}Fig. \ref{fig2}(b){]}, indicating negligible optical
loss and heating effect. This is largely due to the reduced chemical
potential, which compensates the Doppler effect in $I_{2}(2\mu,\mathbf{Q})$
and thereby locks the system in the dark-state regime. Near resonance, the lifetime of
the system is estimated to be $100$ ms \cite{SUPP}.

Numerically we have checked that the FF state is the true minimum
of the energy landscape {[}Fig. \ref{fig3}(a){]} and always has lower
free energy than the $Q=0$ state {[}Fig. \ref{fig3}(b){]}. In the
BEC limit, the FF momentum approaches the momentum of the ground-state
dimer, consistent with the above analysis. Around the MFR, the FF
momentum $Q$ reaches a sizable value $Q\sim k_{{\rm F}}$, which
may lead to visible observational effect in cold atom experiments.
Fig. \ref{fig3}(c) reports a typical energy spectrum of the single-particle
excitation along the ${\bf Q}$ direction, which shows a large anisotropy
between the directions along ${\bf Q}$ and perpendicular to ${\bf Q}$.
The momentum-resolved radio-frequency spectroscopy \cite{Stewart2008}
can be applied to measure this anisotropy and probe the FF state.
The strong violation of Galilean invariance can be seen from the large
difference between the energy gap and pairing gap {[}Fig. \ref{fig3}(b){]}.
As shown in the inset of Fig. \ref{fig3}(c), it also leads to the
significant suppression of superfluid density \cite{Taylor2006} near the resonance at
zero temperature \cite{SUPP}.

We also studied the collective phonon mode, known as the Anderson-Bogoliubov
mode of Fermi superfluidity, by calculating the Gaussian fluctuation
of the effective action around the mean-field solution \cite{SUPP}.
Fig. \ref{fig4} reports the typical behavior of the phonon mode.
Along the ${\bf Q}$ direction, the phonon mode splits into two branches
with different velocities. At large momentum, one branch merges into
the two-particle continuum, leading to an interesting maxon-roton
structure. These predictions can be probed by applying the Bragg spectroscopy
\cite{Veeravalli2008}.

\emph{Summary.} We have proposed that the dark-state optical control
of magnetic Feshbach resonances provides a natural and robust way
to realize the Fulde-Ferrell superfluidity as well as the finite-momentum
BEC of dimers. While our calculations are specific for $^{40}$K atoms,
our theory and mechanism for Fulde-Ferrell superfluidity is generic
and is applicable to other systems including $^{6}$Li atoms. The
unique advantage of our proposal is that the system is free from optical
loss and heating due to the dark-state manipulation. Since no spin-population imbalance is needed,
the Fulde-Ferrell state has a high transition temperature near resonance, which is good for experiments. It opens a fascinating
way to explore some unique features of Fulde-Ferrell superfluids,
in particular, the anisotropic phonon dispersion and emergent roton
structure, by using Bragg spectroscopy.
\begin{acknowledgments}
We thank Professor Peng Zhang for useful discussions. LH was supported
by the Thousand Young Talent Program of China and the National Natural Science Foundation of China (Grant No. 11775123). XJL and HH were supported
under Australian Research Council's Future Fellowships funding scheme
(project No. FT140100003 and No. FT130100815).
\end{acknowledgments}

\pagebreak

\appendix

\begin{widetext}

\section{Supplemental Material}

In this Supplemental Material, we provide detailed information on
the renormalization of coupling parameters, solution of two-body bound
states, analysis of the loss rate of the system, derivation of mean-field
equations, analytic expression in the BEC limit, superfluid density
calculation, and the collective phonon mode.

\subsection{Renormalization and two-body problem}

The atom-molecule theory is a low-energy effective field theory. In
the absence of optical fields, it is designed to recover the known
low-energy atom-atom scattering amplitude $f(k)=1/[k\cot\delta_{s}(k)-ik]$,
with the $s$-wave scattering phase shift given by 
\begin{eqnarray}
k\cot\delta_{s}(k)=-\frac{1}{a_{{\rm bg}}}\frac{E-\Delta\mu(B-B_{0})}{E-\Delta\mu(B-B_{0})+\Delta\mu\Delta B}.\label{AM-Phaseshift}
\end{eqnarray}
Here $E=k^{2}/m$ is the scattering energy in the center-of-mass frame.
In our convention, the resonance width $\Delta B$ can be both positive
and negative, satisfying $\Delta\mu\Delta Ba_{{\rm bg}}>0$. The bare
couplings $g_{0}$ and $u_{0}$ as well as the bare detuning $\nu_{0}$
should be renormalized by the known information of MFR, i.e., the
background scattering length $a_{{\rm bg}}$, the resonance width
$\Delta B$, and the magnetic detuning $\Delta\mu(B-B_{0})$, with
$\Delta\mu$ being the magnetic moment difference between two atoms
and the molecular state $|g_{1}\rangle$. To this end, we compute
the atom-atom scattering amplitude directly from the atom-molecule
theory and match to the known low-energy result (\ref{AM-Phaseshift}).

We first compute the off-shell $T$-matrix for atom-atom scattering,
which is exactly given by the bubble summation, 
\begin{eqnarray}
T_{{\rm 2b}}(q_{0},{\bf q})=\frac{U(q_{0},{\bf q})}{1-U(q_{0},{\bf q})\Pi(q_{0},{\bf q})}=\frac{1}{U(q_{0},{\bf q})^{-1}-\Pi(q_{0},{\bf q})}.\label{AM-Tmatrix}
\end{eqnarray}
Here $q_{0}$ and ${\bf q}$ now stand for the center-of-mass energy
and momentum of the two atoms, respectively. $U(q_{0},{\bf q})$ is
an energy- and momentum-dependent interaction vertex, $U(q_{0},{\bf q})=u_{0}+g_{0}^{2}D_{1}(q_{0},{\bf q})$,
with $D_{1}(q_{0},{\bf q})$ being the propagator of the molecular
state $|g_{1}\rangle$. In the absence of optical fields, $D_{1}(q_{0},{\bf q})$
is given by 
\begin{eqnarray}
D_{1}(q_{0},{\bf q})=\frac{1}{I_{1}(q_{0},{\bf q})}=\frac{1}{Z-\nu_{0}},
\end{eqnarray}
The two-atom bubble function $\Pi(q_{0},{\bf q})$ is given by 
\begin{eqnarray}
\Pi(q_{0},{\bf q})=\sum_{{\bf p}}\frac{1}{Z+i0^{+}-2\varepsilon_{{\bf p}}}.\label{LM-Propagator}
\end{eqnarray}
It is divergent because of the use of contact couplings. We introduce
a large cutoff $\Lambda$ for $|{\bf p}|$ and obtain $\Pi(q_{0},{\bf q})=-\eta(\Lambda)+\Pi_{{\rm R}}(q_{0},{\bf q})$,
with a divergent piece 
\begin{eqnarray}
\eta(\Lambda)=\sum_{{\bf p}}^{|{\bf p}|<\Lambda}\frac{1}{2\varepsilon_{{\bf p}}}=\frac{m\Lambda}{2\pi^{2}}.
\end{eqnarray}
and a finite piece 
\begin{eqnarray}
\Pi_{{\rm R}}(q_{0},{\bf q})=-i\frac{m}{4\pi}\sqrt{m(Z+i0^{+})}.
\end{eqnarray}
In the absence of optical fields, Galiean invariance ensures that
$T_{{\rm 2b}}$ is only a function of $Z=q_{0}-{\bf q}^{2}/(4m)$.
The scattering amplitude $f(k)$ can be obtained by imposing the on-shell
condition $Z=E=k^{2}/m$. We obtain $f(k)=-\frac{m}{4\pi}T_{{\rm 2b}}(Z=k^{2}/m)$.

The renormalization of the atom-molecule theory can be done by matching
the scattering amplitude calculated from the theory with the known
low-energy scattering amplitude (\ref{AM-Phaseshift}). The renormalizability
of the theory requires that the equality 
\begin{eqnarray}
-\eta(\Lambda)-\left[u_{0}(\Lambda)+\frac{g_{0}^{2}(\Lambda)}{E-\nu_{0}(\Lambda)}\right]^{-1}=-\frac{m}{4\pi a_{{\rm bg}}}\frac{E-\Delta\mu(B-B_{0})}{E-\Delta\mu(B-B_{0})+\Delta\mu\Delta B}
\end{eqnarray}
holds for arbitrary value of the scattering energy $E=k^{2}/m$ through
proper cutoff dependence of the bare couplings and the bare detuning.
Defining the renormalized couplings $u=4\pi a_{{\rm bg}}/m$ and $g=\sqrt{\Delta\mu\Delta Bu}$
and the renormalized detuning $\nu=\Delta\mu(B-B_{0})$, we obtain
\cite{Chen2005} 
\begin{eqnarray}
u_{0}(\Lambda)=\frac{u}{1-\eta(\Lambda)u},\ \ \ \ \ \ \ g_{0}(\Lambda)=\frac{g}{1-\eta(\Lambda)u},\ \ \ \ \ \ \nu_{0}(\Lambda)=\nu+\frac{g^{2}\eta(\Lambda)}{1-\eta(\Lambda)u}.
\end{eqnarray}
We also find the following identity, 
\begin{eqnarray}
\left(u+\frac{g^{2}}{X-\nu}\right)^{-1}=\left[u_{0}(\Lambda)+\frac{g_{0}^{2}(\Lambda)}{X-\nu_{0}(\Lambda)}\right]^{-1}+\eta(\Lambda)\label{R-identity}
\end{eqnarray}
holds for arbitrary quantity $X$, which is quite convenient for us
to renormalize the two-body $T$-matrix and the grand potential in
the presence of optical fields.

In the presence of optical fields, the two-body $T$-matrix is given
by 
\begin{eqnarray}
T_{{\rm 2b}}(q_{0},{\bf q})=\frac{1}{U^{-1}(q_{0},{\bf q})-\Pi(q_{0},{\bf q})},
\end{eqnarray}
where 
\begin{eqnarray}
U(q_{0},{\bf q})=u_{0}+\frac{g_{0}^{2}}{Z-\nu_{0}-\Sigma_{1}(q_{0},{\bf q})}.
\end{eqnarray}
Using the result $\Pi(q_{0},{\bf q})=\eta(\Lambda)+\Pi_{{\rm R}}(q_{0},{\bf q})$
and regarding $Z-\Sigma_{1}(q_{0},{\bf q})$ as the quantity $X$
in Eq. (\ref{R-identity}), we obtain the $T$-matrix in terms of
the renormalized quantities, 
\begin{eqnarray}
T_{{\rm 2b}}(q_{0},{\bf q})=\frac{1}{U_{{\rm R}}^{-1}(q_{0},{\bf q})-\Pi_{{\rm R}}(q_{0},{\bf q})},
\end{eqnarray}
where the renormalized effective two-body interaction $U_{{\rm R}}(q_{0},{\bf q})$
is given in the main text. It is evident that the parameters related
to the optical control, i.e., the additional molecular part ${\cal L}_{{\rm M}}^{\prime}$,
does not need renornalization. If there exists a dimer bound state,
its energy $E_{{\rm d}}({\bf q})$ at given center-of-mass momentum
${\bf q}$ is determined by the pole of the $T$-matrix, i.e., 
\begin{eqnarray}
U_{{\rm R}}^{-1}[E_{{\rm d}}({\bf q}),{\bf q}]-\Pi_{{\rm R}}[E_{{\rm d}}({\bf q}),{\bf q}]=0.
\end{eqnarray}
It is evident that the bound-state solution satisfies the condition
$E_{{\rm d}}({\bf q})<{\bf q}^{2}/(4m)$.

\subsection{Decay rate of the dimer bound state }

At the zero relative momentum $\mathbf{k}=0$ and hence $Z=\mathbf{k}^{2}/m=0$
or $q_{0}=\mathbf{q}^{2}/(4m)$, the effective two-body interaction
takes the form (see Eq. (6) in the main text), 
\begin{equation}
U_{\textrm{R}}\left(\mathbf{q}\right)=\frac{4\pi a_{\textrm{bg}}}{m}\left[1+\frac{\Delta\mu\Delta B}{-\Delta\mu\left(B-B_{0}\right)-\Sigma_{1}\left(\mathbf{q}\right)}\right],
\end{equation}
where 
\begin{eqnarray}
\Sigma_{1}\left(\mathbf{q}\right) & = & \frac{\Omega_{1}^{2}/4}{I_{\textrm{e}}(\mathbf{q})-\left(\Omega_{2}^{2}/4\right)/I_{2}\left(\mathbf{q}\right)},\\
I_{\textrm{e}}\left(\mathbf{q}\right) & = & \left(\Delta_{\textrm{e}}+i\frac{\gamma_{\textrm{e}}}{2}\right)-\frac{q_{z}k_{\textrm{R}}}{2m}-\frac{k_{\textrm{R}}^{2}}{4m},\\
I_{2}\left(\mathbf{q}\right) & = & \delta-\frac{q_{z}k_{\textrm{R}}}{m}-\frac{k_{\textrm{R}}^{2}}{m},
\end{eqnarray}
and we already assume $\mathbf{k}_{1}=-\mathbf{k}_{2}=k_{\textrm{R}}\mathbf{e}_{z}$.
Near the resonance with zero two-photon detuning ($\delta=0$), as
the terms $q_{z}k_{\textrm{R}}/m$ and $k_{\textrm{R}}^{2}/m\sim2\pi\times10$
kHz are three orders smaller than $\Omega_{2},I_{\textrm{e}}\sim2\pi\times10$
MHz in magnitude, we may approximate the Stark shift,
\begin{equation}
\Sigma_{1}\left(\mathbf{q}\right)\simeq-\left(\frac{\Omega_{1}}{\Omega_{2}}\right)^{2}I_{2}\left(\mathbf{q}\right)\left[1+\frac{4I_{2}\left(\mathbf{q}\right)}{\Omega_{2}^{2}}I_{\textrm{e}}\left(\mathbf{q}\right)\right].
\end{equation}
Therefore, the effective decay rate $\gamma_{\textrm{eff}}=-\textrm{2Im}\Sigma_{1}(\mathbf{q})$
becomes,
\begin{equation}
\gamma_{\textrm{eff}}=4\left(\frac{\Omega_{1}}{\Omega_{2}}\right)^{2}\left[\frac{I_{2}\left(\mathbf{q}\right)}{\Omega_{2}}\right]^{2}\gamma_{\textrm{e}}.
\end{equation}
By taking the typical values $\Omega_{1}=2\pi\times120$ MHz, $\Omega_{2}=2\pi\times20$
MHz, $I_{2}(\mathbf{q})\sim2\pi\times10$ kHz and $\gamma_{\textrm{e}}=2\pi\times6$
MHz, we find that $\gamma_{\textrm{eff}}\sim3.6\times10^{-5}\gamma_{\textrm{e}}\simeq2\pi\times200$
Hz. On the other hand, the real part of the Stark shift $\textrm{Re}\Sigma_{1}(\mathbf{q})\sim(\Omega_{1}/\Omega_{2})^{2}I_{2}(\mathbf{q})\sim2\pi\times0.4$
MHz. Thus, $\gamma_{\textrm{eff}}$ is three orders smaller than $\textrm{Re}\Sigma_{1}(\mathbf{q})$
in magnitude.

Due to the negligible $\gamma_{\textrm{eff}}$ near the two-photon
resonance, numerically we find that the two-body binding energy of
the dimer bound state and the momentum $Q$ of the dimer are essentially
\emph{independent} on $\gamma_{\textrm{eff}}$ (or $\gamma_{\textrm{e}}$).
For the many-body calculation, we anticipate the results will also
be independent on $\gamma_{\textrm{e}}$. Therefore, for simplicity,
in our mean-field calculations we reasonably set $\gamma_{\textrm{e}}=0$.

Of course, the lifetime of the dimer bound state and Cooper pairs
will depend crucially on the decay rate $\gamma_{\textrm{e}}$, i.e.,
the lifetime will double if we decrease $\gamma_{\textrm{e}}$ by
half. The unique advantage of our dark-state control proposal is that
near the two-photon resonance, the lifetime of these dimers or Cooper
pairs is long enough for the observation of interesting many-body
phenomena such as the Fulde-Ferrell superfluidity. In the next section,
we discuss in detail the two-body collisional loss rate, which should
be taken care of in cooling the Fermi cloud to quantum degeneracy.

\subsection{Two-body loss rate and lifetime of the system }

To calculate the two-body collisional loss rate including the Doppler
and kinetic energy shifts, we must average the loss rate \cite{Jagannathan2016}
\begin{equation}
K_{2}\left(k,\mathbf{q}\right)=\frac{8\pi}{m}\textrm{Im}f\left(k,\mathbf{q}\right)-\frac{8\pi k}{m}\left|f\left(k,\mathbf{q}\right)\right|^{2}
\end{equation}
over the CoM momentum $\mathbf{q}$ and the relative momentum $k$.
Here, the scattering amplitude $f\left(k,\mathbf{q}\right)$ depends
on both $k$ and $\mathbf{q}$ and is given by
\begin{equation}
f\left(k,\mathbf{q}\right)=\frac{1}{-a_{\textrm{R}}^{-1}(k,\mathbf{q})-ik},
\end{equation}
where
\begin{equation}
a_{\textrm{R}}\left(k,\mathbf{q}\right)=\frac{m}{4\pi}U_{\textrm{R}}\left(q_{0}=\frac{k^{2}}{m}+\frac{\mathbf{q}^{2}}{4m},\mathbf{q}\right).
\end{equation}
In cooling the Fermi gas down to the degenerate temperature $T_{F}$,
it is reasonable to assume a classical Boltzmann distribution of the
CoM momentum and the relative momentum \cite{Jagannathan2016}. At
temperature $T$, the momentum averaged loss rate constant then takes
the form,
\begin{equation}
\left\langle K_{2}\right\rangle =\frac{\int d\mathbf{k}d\mathbf{q}K_{2}\left(k,\mathbf{q}\right)\exp\left[-\mathbf{k}^{2}/\left(mk_{B}T\right)\right]\exp\left[-\mathbf{q}^{2}/\left(4mk_{B}T\right)\right]}{\int d\mathbf{k}d\mathbf{q}\exp\left[-\mathbf{k}^{2}/\left(mk_{B}T\right)\right]\exp\left[-\mathbf{q}^{2}/\left(4mk_{B}T\right)\right]}.
\end{equation}
By noting that $K_{2}\left(k,\mathbf{q}\right)$ depends on $q_{z}$
only, we have the expression,
\begin{equation}
\left\langle K_{2}\right\rangle =\frac{\int_{0}^{\infty}k^{2}dk\int_{-\infty}^{\infty}dq_{z}K_{2}\left(k,q_{z}\right)\exp\left[-k^{2}/\left(mk_{B}T\right)\right]\exp\left[-q_{z}^{2}/\left(4mk_{B}T\right)\right]}{\int_{0}^{\infty}k^{2}dk\int_{-\infty}^{\infty}dq_{z}\exp\left[-k^{2}/\left(mk_{B}T\right)\right]\exp\left[-q_{z}^{2}/\left(4mk_{B}T\right)\right]}.
\end{equation}
To perform the numerical calculation, we introduce $\tilde{a}_{\textrm{R}}=a_{\textrm{R}}/a_{\textrm{bg}}$
and take $k_{\textrm{R}}$ and $E_{\textrm{R}}$ as the units for
momentum and energy/temperature (i.e., $\tilde{k}=k/k_{\textrm{R}}$,
$\tilde{q}_{z}=q_{z}/k_{\textrm{R}}$ and $\tilde{T}=k_{B}T/E_{\rm R}$
are to be used), respectively. Thus, we have,
\begin{equation}
\left\langle K_{2}\right\rangle =2\left(\frac{8}{3\pi}\right)\left(\frac{E_{\textrm{R}}}{n}\right)\left(k_{\textrm{R}}a_{\textrm{bg}}\right)\left(\frac{8}{\pi\tilde{T}^{2}}\right)\int_{0}^{\infty}\tilde{k}^{2}d\tilde{k}\int_{-\infty}^{\infty}d\tilde{q}_{z}\tilde{K}_{2}\left(\tilde{k},\tilde{q}_{z}\right)\exp\left[-2\tilde{k}^{2}/\tilde{T}\right]\exp\left[-\tilde{q}_{z}^{2}/\left(2\tilde{T}\right)\right],
\end{equation}
where the characteristic density $n=k_{\textrm{F}}^{3}/(3\pi^{2})=k_{\textrm{R}}^{3}/(3\pi^{2})$
and we have defined,
\begin{equation}
\tilde{K}_{2}\left(\tilde{k},\tilde{q}_{z}\right)=\frac{\textrm{Im}\tilde{a}_{\rm R}^{-1}\left(\tilde{k},\tilde{q}_{z}\right)}{\left[\textrm{Re}\tilde{a}_{\rm R}^{-1}\left(\tilde{k},\tilde{q}_{z}\right)\right]^{2}+\left[\left(k_{\textrm{R}}a_{\textrm{bg}}\right)\tilde{k}+\textrm{Im}\tilde{a}_{\rm R}^{-1}\left(\tilde{k},\tilde{q}_{z}\right)\right]^{2}},
\end{equation}
and
\begin{eqnarray}
\tilde{a}_{\rm R}\left(\tilde{k},\tilde{q}_{z}\right) & = & 1+\frac{\Delta\mu\Delta B/E_{\textrm{R}}}{2\tilde{k}^{2}-\Delta\mu\left(B-B_{0}\right)/E_{\textrm{R}}-\left[\Omega_{1}^{2}/\left(4E_{\textrm{R}}^{2}\right)\right]\left[\tilde{I}_{\textrm{e}}(\tilde{k},\tilde{q}_{z})-\left(\Omega_{2}^{2}/4E_{\textrm{R}}^{2}\right)/\tilde{I}_{\textrm{2}}(\tilde{k},\tilde{q}_{z})\right]^{-1}},\\
\tilde{I}_{\textrm{e}}(\tilde{k},\tilde{q}_{z}) & = & \left(\Delta_{\textrm{e}}+i\frac{\gamma_{\textrm{e}}}{2}\right)/E_{\textrm{R}}+2\tilde{k}^{2}-\tilde{q}_{z}-1/2,\\
\tilde{I}_{2}(\tilde{k},\tilde{q}_{z}) & = & \delta/E_{\textrm{R}}+2\tilde{k}^{2}-2\tilde{q}_{z}-2.
\end{eqnarray}
Once the averaged $\left\langle K_{2}\right\rangle $ constant is
obtained, we calculate the lifetime of the system by using,
\begin{equation}
\tau_{\textrm{loss}}=\frac{2}{n\left\langle K_{2}\right\rangle }.
\end{equation}
Here the factor of $2$ comes from the fact that the density of each
spin-population is $n/2$.

\begin{figure}[h]
\centering{}\includegraphics[width=0.9\textwidth]{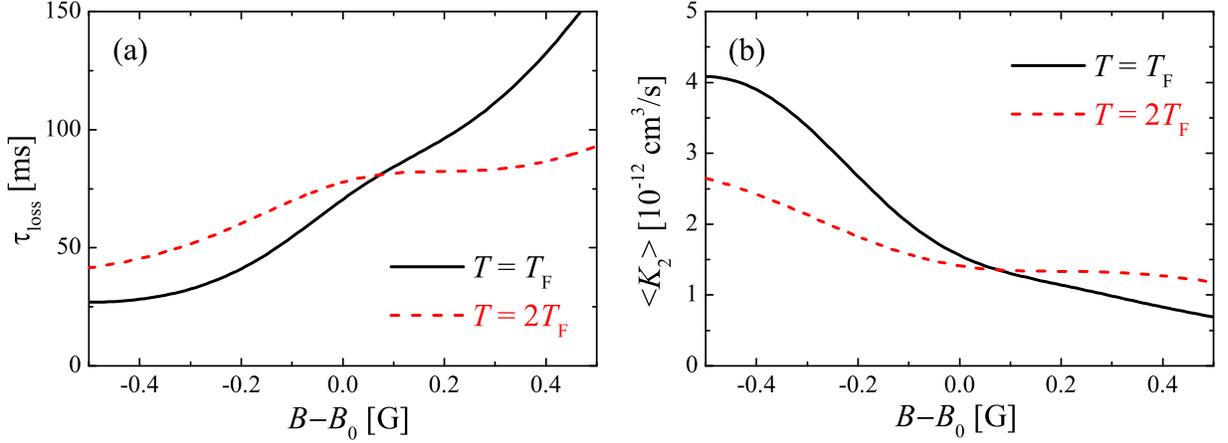} \caption{The lifetime (a) and the averaged $K_{2}$ constant (b) near the Feshbach
resonance, including the Doppler and kinetic energy shifts, at two
temperatures $T=T_{\rm F}$ (solid line) and $T=2T_{\rm F}$ (dashed line).
Here, we take the typical experimental parameters detailed in the
text. \label{fig:lifetime}}
\end{figure}

By considering a Fermi gas of $^{40}$K atoms at the broad Feshbach
resonance $B_{0}=202.02$ G and taking $k_{\textrm{R}}=k_{\textrm{F}}=8.138\times10^{6}$
m$^{-1}$ (which corresponds to an atom density $n\simeq1.82\times10^{13}$
cm$^{-3}$ and $E_{\textrm{F}}=E_{\textrm{R}}\simeq2\pi\times8.36$
kHz) as described in the main text, we find that $k_{\textrm{R}}a_{\textrm{bg}}\simeq0.07491$
and $[3\pi/(8E_{\textrm{R}})](k_{\textrm{R}}a_{\textrm{bg}})^{-1}\simeq0.30$
ms, and then,
\begin{equation}
\tau_{\textrm{loss}}\simeq\left(\frac{\pi\tilde{T}^{2}}{8}\right)\frac{0.30\textrm{ ms}}{\int_{0}^{\infty}\tilde{k}^{2}d\tilde{k}\int_{-\infty}^{\infty}d\tilde{q}_{z}\tilde{K}_{2}\left(\tilde{k},\tilde{q}_{z}\right)\exp\left[-2\tilde{k}^{2}/\tilde{T}\right]\exp\left[-\tilde{q}_{z}^{2}/\left(2\tilde{T}\right)\right]}.
\end{equation}
The integral can be easily calculated. Using the typical values for
the dark-state control as listed in the main text, we obtain the lifetime
of the system near the Feshbach resonance at two temperatures $T=T_{\rm F}\sim0.4$
$\mu$K (solid line) and $T=2T_{\rm F}\sim0.8$ $\mu$K (dashed line),
as reported in Fig. \ref{fig:lifetime}(a). We find that the lifetime
is about $\sim50$ ms. Thus, near the Feshbach resonance, the lifetime
of the dark-state controlled Fermi gas can be enhanced to the same
order in magnitude as that of a Fermi gas \emph{without} optical control.
In the latter case, the lifetime of the system ($\sim100$ ms as reported
in Ref. \cite{Regal2004}) is limited by three-body recombination
process for dimers or spin-flip for atoms, and the reach of fermionic
superfluidity at the BEC-BCS crossover has been routinely demonstrated
in cold-atom laboratories. The corresponding averaged $\left\langle K_{2}\right\rangle $
constant is shown in Fig. \ref{fig:lifetime}(b). It is about $10^{-12}$
cm$^{3}$/s, slightly above the Feshbach resonance.

\subsection{Mean-field equations}

The saddle-point solutions for the Fulde-Ferrell state take the form
\begin{eqnarray}
\bar{\phi}_{{\rm f}}({\bf r})=C_{{\rm f}}e^{i{\bf Q}\cdot{\bf r}},\ \ \ \ \ \bar{\phi}_{1}({\bf r})=C_{1}e^{i{\bf Q}\cdot{\bf r}},\ \ \ \ \ \ \Delta({\bf r})=\Delta e^{i{\bf Q}\cdot{\bf r}},
\end{eqnarray}
where ${\bf Q}$ is the pairing momentum and $\Delta=C_{{\rm f}}+g_{0}C_{1}$.
From the saddle point condition, we find that the self-consistent
solutions of $\bar{\phi}_{2}({\bf r})$ and $\bar{\phi}_{{\rm e}}({\bf r})$
take the same form, i.e., $\bar{\phi}_{2}({\bf r})=C_{2}e^{i{\bf Q}\cdot{\bf r}}$
and $\bar{\phi}_{{\rm e}}({\bf r})=C_{{\rm e}}e^{i{\bf Q}\cdot{\bf r}}$.
Then we can evaluate the effective action in terms of $C_{l}$ and
${\bf Q}$. The grand potential at $T=0$ in mean-field approximation
reads 
\begin{eqnarray}
\Omega=\Omega_{{\rm q}}+\sum_{{\bf k}}(\xi_{{\bf k}}-E_{{\bf k}})-C^{\dagger}{\bf M}(2\mu,{\bf Q})C-\frac{|C_{{\rm f}}|^{2}}{u_{0}}.\label{Omega}
\end{eqnarray}
where the quasiparticle contribution is given by 
\begin{eqnarray}
\Omega_{{\rm q}}=\sum_{s=\pm}\sum_{{\bf k}}E_{{\bf k}}^{s}\Theta(-E_{{\bf k}}^{s})=2\sum_{{\bf k}}E_{{\bf k}}^{+}\Theta(-E_{{\bf k}}^{+}).
\end{eqnarray}
Here the vector $C$ is defined as $C=(C_{1},\ C_{2},\ C_{{\rm e}})^{{\rm T}}$.
It is evident that $\Omega_{{\rm q}}$ contributes only if the quasiparticle
excitations $E_{{\bf k}}^{\pm}=E_{{\bf k}}\pm{\bf k}\cdot{\bf Q}/(2m)$
are gapless, i.e., $\mu>0$ and $|{\bf Q}|>(|\Delta|/\mu)\sqrt{2m\mu}$.

The second term of the expression (\ref{Omega}) is divergent. Note
that it contains bare quantities, i.e., $u_{0}$, $g_{0}$, $\nu_{0}$,
and $C_{l}$ ($l=1,2,{\rm e},{\rm f}$). Using the saddle-point condition
$\partial\Omega/\partial C_{2}=0$ and $\partial\Omega/\partial C_{{\rm e}}=0$,
we can express $C_{2}$ and $C_{{\rm e}}$ in terms of $C_{1}$. Then
eliminating $C_{2}$ and $C_{{\rm e}}$, we obtain 
\begin{eqnarray}
\Omega=\Omega_{{\rm q}}+\sum_{{\bf k}}(\xi_{{\bf k}}-E_{{\bf k}})-\frac{|C_{{\rm f}}|^{2}}{u_{0}}-\left[2\mu-\frac{{\bf Q}^{2}}{4m}-\nu_{0}-\Sigma_{1}(2\mu,{\bf Q})\right]|C_{1}|^{2}.
\end{eqnarray}
Again using the saddle-point condition $\partial\Omega/\partial C_{1}=0$
and $\partial\Omega/\partial C_{{\rm f}}=0$, we can express $C_{1}$
and $C_{{\rm f}}$ in terms of the physical quantity $\Delta=C_{{\rm f}}+g_{0}C_{1}$.
Then eliminating $C_{1}$ and $C_{{\rm f}}$, and using the identity
(\ref{R-identity}), we finally obtain 
\begin{eqnarray}
\Omega=\Omega_{{\rm q}}+\sum_{{\bf k}}\left(\xi_{{\bf k}}-E_{{\bf k}}+\frac{|\Delta|^{2}}{2\varepsilon_{{\bf k}}}\right)-\frac{|\Delta|^{2}}{U_{{\rm R}}(2\mu,{\bf Q})},\label{R-Omega}
\end{eqnarray}
which is free from the ultraviolet cutoff. It is obvious that $\Delta$
can be set to be real without loss of generality.

We consider two counterpropagating optical fields, ${\bf k}_{1}=-{\bf k}_{2}=k_{{\rm R}}{\bf e}_{z}$,
and take ${\bf Q}=Q{\bf e}_{z}$. Completing the angle integration
in Eq. (\ref{R-Omega}), we obtain 
\begin{eqnarray}
\Omega=\int_{0}^{\infty}\frac{k^{2}dk}{2\pi^{2}}\left(\xi_{k}-E_{k}+\frac{\Delta^{2}}{2\varepsilon_{k}}\right)-\frac{m}{|Q|}\int_{0}^{\infty}\frac{kdk}{2\pi^{2}}\left(E_{k}-\frac{k|Q|}{2m}\right)^{2}
\Theta\left(\frac{k|Q|}{2m}-E_{k}\right)-\frac{\Delta^{2}}{U_{{\rm R}}(2\mu,Q)},
\end{eqnarray}
where 
\begin{eqnarray}
U_{{\rm R}}(2\mu,Q) & = & u+g^{2}\left[2\mu-\frac{Q^{2}}{4m}-\nu-\Sigma_{1}(2\mu,Q)\right]^{-1},\nonumber \\
\Sigma_{1}(2\mu,Q) & = & \frac{\Omega_{1}^{2}}{4}\left[I_{{\rm e}}(2\mu,Q)-\frac{\Omega_{2}^{2}}{4I_{2}(2\mu,Q)}\right]^{-1}.
\end{eqnarray}
Here 
\begin{eqnarray}
 &  & I_{{\rm e}}(2\mu,Q)=2\mu-\frac{Q^{2}}{4m}-\frac{E_{{\rm R}}}{2}-\frac{k_{{\rm R}}Q}{2m}+\Delta_{{\rm e}},\nonumber \\
 &  & I_{2}(2\mu,Q)=2\mu-\frac{Q^{2}}{4m}-2E_{{\rm R}}-\frac{k_{{\rm R}}Q}{m}-\delta.
\end{eqnarray}

The gap equation $\partial\Omega/\partial\Delta$ reads 
\begin{eqnarray}
-\frac{1}{U_{{\rm R}}(2\mu,Q)}=\int_{0}^{\infty}\frac{k^{2}dk}{2\pi^{2}}\left(\frac{1}{2E_{k}}-\frac{1}{2\varepsilon_{k}}\right)+\frac{m}{|Q|}\int_{0}^{\infty}\frac{kdk}{2\pi^{2}}\left(1-\frac{k|Q|}{2mE_{k}}\right)
\Theta\left(\frac{k|Q|}{2m}-E_{k}\right).
\end{eqnarray}
The number equation $n=-\partial\Omega/\partial\mu$ can be evaluated
as 
\begin{eqnarray}
n & = & \int_{0}^{\infty}\frac{k^{2}dk}{2\pi^{2}}\left(1-\frac{\xi_{k}}{E_{k}}\right)-\frac{2m}{|Q|}\int_{0}^{\infty}\frac{kdk}{2\pi^{2}}\frac{\xi_{k}}{E_{k}}\left(E_{k}-\frac{k|Q|}{2m}\right)\Theta\left(\frac{k|Q|}{2m}-E_{k}\right)\nonumber \\
 & + & \frac{2\Delta^{2}}{g^{2}}\left[1-\frac{u}{U_{{\rm R}}(2\mu,Q)}\right]^{2}\left\{ 1+\frac{\frac{1}{4}\Omega_{1}^{2}\left[I_{2}^{2}(2\mu,Q)+\frac{1}{4}\Omega_{2}^{2}\right]}{\left[I_{{\rm e}}(2\mu,Q)I_{2}(2\mu,Q)-\frac{1}{4}\Omega_{2}^{2}\right]^{2}}\right\} .
\end{eqnarray}
Meanwhile, we show that 
\begin{eqnarray}
\frac{\partial\Omega}{\partial Q}=W_{1}(Q)+W_{2}(Q),
\end{eqnarray}
where 
\begin{eqnarray}
W_{1}(Q) & = & \frac{Q}{4m}\int_{0}^{\infty}\frac{k^{2}dk}{2\pi^{2}}\left(1-\frac{\xi_{k}}{E_{k}}\right)+{\rm sgn}(Q)\frac{m}{|Q|^{2}}\int_{0}^{\infty}\frac{kdk}{2\pi^{2}}\left(E_{k}-\frac{k|Q|}{2m}\right)^{2}
\Theta\left(\frac{k|Q|}{2m}-E_{k}\right)\nonumber \\
 & - & {\rm sgn}(Q)\frac{m}{|Q|}\int_{0}^{\infty}\frac{kdk}{2\pi^{2}}\left(E_{k}-\frac{k|Q|}{2m}\right)\left(\frac{\xi_{k}}{E_{k}}\frac{|Q|}{2m}-\frac{k}{m}\right)\Theta\left(\frac{k|Q|}{2m}-E_{k}\right)\nonumber \\
 & + & \frac{Q}{2m}\frac{\Delta^{2}}{g^{2}}\left[1-\frac{u}{U_{{\rm R}}(2\mu,Q)}\right]^{2}\left\{ 1+\frac{\frac{1}{4}\Omega_{1}^{2}\left[I_{2}^{2}(2\mu,Q)+\frac{1}{4}\Omega_{2}^{2}\right]}{\left[I_{{\rm e}}(2\mu,Q)I_{2}(2\mu,Q)-\frac{1}{4}\Omega_{2}^{2}\right]^{2}}\right\} ,\nonumber \\
W_{2}(Q) & = & \frac{k_{{\rm R}}}{2m}\frac{\Delta^{2}}{g^{2}}\left[1-\frac{u}{U_{{\rm R}}(2\mu,Q)}\right]^{2}\frac{\frac{1}{4}\Omega_{1}^{2}\left[\frac{1}{2}\Omega_{2}^{2}+I_{2}^{2}(2\mu,Q)\right]}{\left[I_{{\rm e}}(2\mu,Q)I_{2}(2\mu,Q)-\frac{1}{4}\Omega_{2}^{2}\right]^{2}}.
\end{eqnarray}
At $Q=0$, we have $W_{1}(Q)=0$ but $W_{2}(Q)\neq0$. Therefore,
$Q=0$ does not satisfy $\partial\Omega/\partial Q=0$.

Using the above results, we also obtain the current ${\bf j}$ in
the zero-momentum pairing state (${\bf Q}=0$). We have 
\begin{eqnarray}
{\bf j}=2m\frac{\partial\Omega}{\partial{\bf Q}}\bigg|_{{\bf Q=0}}.
\end{eqnarray}
It is obvious that $j_{x}=j_{y}=0$, and the current along the $z$-direction
reads 
\begin{eqnarray}
j_{z}=k_{{\rm R}}\frac{\Delta^{2}}{g^{2}}\left[1-\frac{u}{U_{{\rm R}}(2\mu,0)}\right]^{2}\frac{\frac{1}{4}\Omega_{1}^{2}\left[\frac{1}{2}\Omega_{2}^{2}+I_{2}^{2}(2\mu,0)\right]}{\left[I_{{\rm e}}(2\mu,0)I_{2}(2\mu,0)-\frac{1}{4}\Omega_{2}^{2}\right]^{2}}.
\end{eqnarray}

\subsection{BEC limit}

In the BEC limit, we have $\tilde{\mu}=\mu-Q^{2}/(8m)<0$ and $|\tilde{\mu}|\gg\Delta$.
To the leading order in $\Delta/|\tilde{\mu}|$, the gap equation
becomes 
\begin{eqnarray}
J(\mu,Q)=0,
\end{eqnarray}
where 
\begin{eqnarray}
J(\mu,Q) & = & -\frac{1}{U_{{\rm R}}(2\mu,Q)}+\int_{0}^{\infty}\frac{k^{2}dk}{2\pi^{2}}\left(\frac{1}{2\varepsilon_{k}}-\frac{1}{2\varepsilon_{k}-2\tilde{\mu}}\right)\nonumber \\
 & = & -\frac{1}{U_{{\rm R}}(2\mu,Q)}+\frac{m}{4\pi}\sqrt{-m\left(2\mu-\frac{Q^{2}}{4m}\right)}.
\end{eqnarray}
Note that we have used the fact that the quasiparticles are gapped.
The number equation becomes 
\begin{eqnarray}
n=2\Delta^{2}\int_{0}^{\infty}\frac{k^{2}dk}{2\pi^{2}}\frac{1}{\left(2\varepsilon_{k}-2\tilde{\mu}\right)^{2}}+\frac{2\Delta^{2}}{g^{2}}\left[1-\frac{u}{U_{{\rm R}}(2\mu,Q)}\right]^{2}\left\{ 1+\frac{\frac{1}{4}\Omega_{1}^{2}\left[I_{2}^{2}(2\mu,Q)+\frac{1}{4}\Omega_{2}^{2}\right]}{\left[I_{{\rm e}}(2\mu,Q)I_{2}(2\mu,Q)-\frac{1}{4}\Omega_{2}^{2}\right]^{2}}\right\} .
\end{eqnarray}
Using this result, we find that 
\begin{eqnarray}
\frac{\partial\Omega}{\partial Q}=\frac{Q}{4m}n+\frac{k_{{\rm R}}}{2m}\frac{\Delta^{2}}{g^{2}}\left[1-\frac{u}{U_{{\rm R}}(2\mu,Q)}\right]^{2}\frac{\frac{1}{4}\Omega_{1}^{2}\left[\frac{1}{2}\Omega_{2}^{2}+I_{2}^{2}(2\mu,Q)\right]}{\left[I_{{\rm e}}(2\mu,Q)I_{2}(2\mu,Q)-\frac{1}{4}\Omega_{2}^{2}\right]^{2}}.
\end{eqnarray}

The equation $J(\mu,Q)=0$ determines the chemical potential as a
function of $Q$, i.e., $\mu=\mu(Q)$. Then we obtain 
\begin{eqnarray}
\frac{\partial J(\mu,Q)}{\partial Q}+\frac{\partial J(\mu,Q)}{\partial\mu}\frac{\partial\mu(Q)}{\partial Q}=0.
\end{eqnarray}
Meanwhile, the grand potential can be expressed as 
\begin{eqnarray}
\Omega=J(\mu,Q)\Delta^{2}+O(\Delta^{4}),
\end{eqnarray}
which leads to 
\begin{eqnarray}
\frac{\partial J(\mu,Q)}{\partial Q}=\frac{1}{\Delta^{2}}\frac{\partial\Omega}{\partial Q},\ \ \ \ \ \ \ \ \frac{\partial J(\mu,Q)}{\partial\mu}=\frac{1}{\Delta^{2}}\frac{\partial\Omega}{\partial\mu}=-\frac{n}{\Delta^{2}}.
\end{eqnarray}
Using the fact $\partial\Omega/\partial Q=0$, we obtain in the BEC
limit 
\begin{eqnarray}
\frac{\partial\mu(Q)}{\partial Q}=0.
\end{eqnarray}

\subsection{Superfluid density}

The superfluid density can be conveniently calculated by using the
phase-twist method, i.e., adding a small boost 
\begin{equation}
{\bf v}_{s}=\frac{{\bf Q}_{s}}{2m}
\end{equation}
to the system \cite{Taylor2006}. The condensates transform like $\bar{\phi}_{l}({\bf r})\rightarrow\bar{\phi}_{l}({\bf r})e^{i{\bf Q}_{s}\cdot{\bf r}}$.
The response of the system at a given chemical potential gives the
superfluid density tensor ($i,j=x,y,z$) 
\begin{equation}
n_{s,ij}=\frac{1}{m}\frac{d^{2}\Omega\left({\bf v}_{s}\right)}{dv_{s,i}dv_{s,j}}\bigg|_{{\bf v}_{s}=0}=4m\frac{d^{2}\Omega\left({\bf Q}_{s}\right)}{dQ_{s,i}dQ_{s,j}}\bigg|_{{\bf Q}_{s}=0}.
\end{equation}
For the FF state, we have $n_{s,ij}=0$ for $i\neq j$ and $n_{s,xx}=n_{s,yy}\neq n_{s,zz}$.
It is evident that at $T=0$, $n_{s,xx}=n_{s,yy}=n$, since the quasipaticles
are gapped and the Galilean invariance is preserved on the $x-y$
plane. Let us consider the superfluid density along the FF momentum
${\bf Q}=Q{\bf e}_{z}$ (i.e., ${\bf Q}_{s}//{\bf Q}$). For a {\em
fixed} chemical potential, we vary the momentum $Q$ and solve the
pairing gap $\Delta\left(Q\right)$ by using 
\begin{equation}
\frac{\partial\Omega\left[Q,\Delta\left(Q\right)\right]}{\partial\Delta}=0,
\end{equation}
which leads to 
\begin{equation}
\frac{\partial^{2}\Omega}{\partial\Delta^{2}}\frac{d\Delta}{dQ}+\frac{\partial^{2}\Omega}{\partial\Delta\partial Q}=0
\end{equation}
or 
\begin{equation}
\frac{d\Delta}{dQ}=-\frac{\partial^{2}\Omega}{\partial\Delta\partial Q}\left(\frac{\partial^{2}\Omega}{\partial\Delta^{2}}\right)^{-1}.\label{dDeltadQ}
\end{equation}
Meanwhile, we have 
\begin{equation}
\frac{d\Omega}{dQ}=\frac{\partial\Omega}{\partial Q}+\frac{\partial\Omega}{\partial\Delta}\frac{d\Delta}{dQ}
\end{equation}
and 
\begin{equation}
\frac{d^{2}\Omega}{dQ^{2}}=\frac{\partial^{2}\Omega}{\partial Q^{2}}+2\frac{\partial^{2}\Omega}{\partial\Delta\partial Q}\frac{d\Delta}{dQ}+\frac{\partial^{2}\Omega}{\partial\Delta^{2}}\left(\frac{d\Delta}{dQ}\right)^{2}.
\end{equation}
By using Eq. (\ref{dDeltadQ}), we obtain 
\begin{equation}
n_{s,zz}=4m\left[\frac{\partial^{2}\Omega}{\partial Q^{2}}-\left(\frac{\partial^{2}\Omega}{\partial\Delta\partial Q}\right)^{2}\left(\frac{\partial^{2}\Omega}{\partial\Delta^{2}}\right)^{-1}\right],
\end{equation}
where all the second derivatives are calculated with the mean-field
solution ($\mu,\Delta,Q$). It is evident that $n_{s,zz}\neq n$ for
the FF state even at $T=0$.

\subsection{Collective Phonon mode}

The collective modes can be investigated by computing the effective
action from the Gaussian fluctuations around the mean field \cite{Ohashi2003}.
The detailed derivation of the effective action will be presented
in a long sequent paper. The effective action for the collective phonon
mode, or the so-called Anderson-Bogoliubov mode of Fermi superfluidity,
is given by 
\begin{eqnarray}
{\cal S}_{{\rm coll}}=\frac{1}{2}\sum_{q}\left(\delta\Delta_{q}^{*}\ \ \delta\Delta_{-q}\right)\left[\begin{array}{ll}
M_{11}(q) & M_{12}(q)\\
M_{21}(q) & M_{22}(q)
\end{array}\right]\left(\begin{array}{l}
\delta\Delta_{q}\\
\delta\Delta_{-q}^{*}
\end{array}\right),
\end{eqnarray}
where we write $\Delta(x)=\Delta+\delta\Delta(x)$ with $\delta\Delta(x)$
being the quantum fluctuation around the mean field $\Delta$, and
$\delta\Delta_{q}$ is the Fourier component of $\delta\Delta(x)$.
Here $q=(i\nu_{n},{\bf q})$ with $\nu_{n}=2\pi nT$ being the boson
Matsubara frequency. The inverse propagator matrix $M(q)$ determines
the properties the collective modes. Its elements satisfies $M_{22}(q)=M_{11}(-q)$
and $M_{21}(q)=M_{12}(-q)$. The explicit form of $M_{11}(q)$ can
be evaluated as 
\begin{eqnarray}
M_{11}(i\nu_{n},{\bf q}) & = & -\frac{1}{U_{{\rm R}}\left(i\nu_{n}+2\mu,{\bf q}+{\bf Q}\right)}+\sum_{{\bf k}}\left[\frac{1}{2\varepsilon_{{\bf k}}}+u_{{\bf k}+{\bf q}/2}^{2}u_{{\bf k}-{\bf q}/2}^{2}\frac{1-f_{{\bf k}+{\bf q}/2}^{(+)}-f_{{\bf k}-{\bf q}/2}^{(-)}}{i\nu_{n}-{\bf q\cdot Q}/(2m)-E_{{\bf k}+{\bf q}/2}-E_{{\bf k}-{\bf q}/2}}\right.\nonumber \\
 &  & -\left.u_{{\bf k}+{\bf q}/2}^{2}v_{{\bf k}-{\bf q}/2}^{2}\frac{f_{{\bf k}+{\bf q}/2}^{(+)}-f_{{\bf k}-{\bf q}/2}^{(+)}}{i\nu_{n}-{\bf q\cdot Q}/(2m)-E_{{\bf k}+{\bf q}/2}+E_{{\bf k}-{\bf q}/2}}\right.\nonumber \\
 &  & +\left.v_{{\bf k}+{\bf q}/2}^{2}u_{{\bf k}-{\bf q}/2}^{2}\frac{f_{{\bf k}+{\bf q}/2}^{(-)}-f_{{\bf k}-{\bf q}/2}^{(-)}}{i\nu_{n}-{\bf q\cdot Q}/(2m)+E_{{\bf k}+{\bf q}/2}-E_{{\bf k}-{\bf q}/2}}\right.\nonumber \\
 &  & -\left.v_{{\bf k}+{\bf q}/2}^{2}v_{{\bf k}-{\bf q}/2}^{2}\frac{1-f_{{\bf k}+{\bf q}/2}^{(-)}-f_{{\bf k}-{\bf q}/2}^{(+)}}{i\nu_{n}-{\bf q\cdot Q}/(2m)+E_{{\bf k}+{\bf q}/2}+E_{{\bf k}-{\bf q}/2}}\right]
\end{eqnarray}
where $f_{{\bf k}}^{(\pm)}=f\left(E_{{\bf k}}^{\pm}\right)$ with
$f(x)=1/(e^{x/T}+1)$ being the Fermi-Dirac distribution. Here the
BCS distributions are defined as $u_{{\bf k}}^{2}=(1+\xi_{{\bf k}}/E_{{\bf k}})/2$
and $v_{{\bf k}}^{2}=1-u_{{\bf k}}^{2}$. The zero-temperature result
is obtained by taking the limit $T\rightarrow0$. The explicit form
of $M_{12}(q)$ reads 
\begin{eqnarray}
M_{12}(i\nu_{n},{\bf q}) & = & \sum_{{\bf k}}\left(uv\right)_{{\bf k}+{\bf q}/2}\left(uv\right)_{{\bf k}-{\bf q}/2}\left[-\frac{1-f_{{\bf k}+{\bf q}/2}^{(+)}-f_{{\bf k}-{\bf q}/2}^{(-)}}{i\nu_{n}-{\bf q\cdot Q}/(2m)-E_{{\bf k}+{\bf q}/2}-E_{{\bf k}-{\bf q}/2}}\right.\nonumber \\
 &  & -\left.\frac{f_{{\bf k}+{\bf q}/2}^{(+)}-f_{{\bf k}-{\bf q}/2}^{(+)}}{i\nu_{n}-{\bf q\cdot Q}/(2m)-E_{{\bf k}+{\bf q}/2}+E_{{\bf k}-{\bf q}/2}}\right.\nonumber \\
 &  & +\left.\frac{f_{{\bf k}+{\bf q}/2}^{(-)}-f_{{\bf k}-{\bf q}/2}^{(-)}}{i\nu_{n}-{\bf q\cdot Q}/(2m)+E_{{\bf k}+{\bf q}/2}-E_{{\bf k}-{\bf q}/2}}\right.\nonumber \\
 &  & +\left.\frac{1-f_{{\bf k}+{\bf q}/2}^{(-)}-f_{{\bf k}-{\bf q}/2}^{(+)}}{i\nu_{n}-{\bf q\cdot Q}/(2m)+E_{{\bf k}+{\bf q}/2}+E_{{\bf k}-{\bf q}/2}}\right].
\end{eqnarray}

The dispersion relation $\omega\left({\bf q}\right)$ of the phonon
mode is determined by 
\begin{equation}
M_{11}\left(\omega,{\bf q}\right)M_{22}\left(\omega,{\bf q}\right)-M_{12}\left(\omega,{\bf q}\right)M_{21}\left(\omega,{\bf q}\right)=0.
\end{equation}
We can show that the above equation holds exactly for $(\omega,{\bf q})=(0,{\bf 0})$.
Naïvely, we may anticipate that in the long-wavelength limit, 
\begin{equation}
\omega\left({\bf q}\right)\approx cq+\frac{{\bf q\cdot Q}}{2m}.
\end{equation}
Therefore, if ${\bf q//Q}$, we would have two branches of phonon
modes 
\begin{eqnarray}
\omega_{+}\left({\bf q}\right)\approx\left(c+\frac{Q}{2m}\right)q,\ \ \ \ \ \ \omega_{-}\left({\bf q}\right)\approx\left(c-\frac{Q}{2m}\right)q.
\end{eqnarray}

\end{widetext}


\begin{thebibliography}{10}
\bibitem{Note-gammae} {We have solved the two-body problem with and without the decay rate $\gamma_{\rm e}$. The difference in results is extremely small and is about $10^{-3}$ in relative, due to the smallness of the effective decay rate. We therefore conclude that the interesting physics is essentially unaffected by the decay rate $\gamma_{\rm e}$. In solving the many-body problem, for simplicity we 
have set  $\gamma_{\rm e}=0$ in mean-field calculations. The resulting many-body chemical potential and FF momentum are then 
used to determine the many-body Stark shfit.}
\end{thebibliography}
\end{document}